\documentclass[11pt,a4paper,superscriptaddress,prb,amsmath,amssymb,aps,letterpaper,final,showkeys]{revtex4}
\usepackage{graphicx,graphicx,epsfig,amsmath,mathtools}

\usepackage{graphicx}
\usepackage{bm}

\newcommand{\la}{\langle} 
\newcommand{\ra}{\rangle}

\begin{document}
\title{Coarse-graining stochastic biochemical networks:
quasi-stationary approximation and fast simulations using a
stochastic path integral technique}

\author{N.A. Sinitsyn} 
\email{nsinitsyn@lanl.gov}
\affiliation{Center for Nonlinear Studies and Computer, Computational
and Statistical Sciences Division, Los Alamos National Laboratory,
Los Alamos, NM 87545 USA}

\author{Nicolas Hengartner} \email{nickh@lanl.gov}
\affiliation{Computer, Computational and Statistical Sciences
  Division, Los Alamos National Laboratory, Los Alamos, NM 87545 USA}

\author{Ilya Nemenman} \email{nemenman@lanl.gov} \affiliation{Center
 for Nonlinear Studies and Computer, Computational and Statistical
 Sciences Division, Los Alamos National Laboratory, Los Alamos, NM
 87545 USA}


\begin{abstract}
  We propose a universal approach for analysis and fast simulations of
  stiff stochastic biochemical kinetics networks, which rests on
  elimination of fast chemical species without a loss of information
  about mesoscopic, non-Poissonian fluctuations of the slow ones.  Our
  approach, which is similar to the Born-Oppenheimer approximation in
  quantum mechanics, follows from the stochastic path integral
  representation of the full counting statistics of reaction events
  (also known as the cumulant generating function). In applications
  with a small number of chemical reactions, this approach produces
  analytical expressions for moments of chemical fluxes between slow
  variables. This allows for a low-dimensional, interpretable
  representation of the biochemical system, that can be used for
  coarse-grained numerical simulation schemes with a small
  computational complexity and yet high accuracy.  As an example, we
  consider a chain of biochemical reactions, derive its coarse-grained
  description, and show that the Gillespie simulations of the original
  stiff system, the coarse-grained simulations, and the full
  analytical treatment are in an agreement, but the coarse-grained
  simulations are three orders of magnitude faster than the Gillespie
  analogue.
\end{abstract}


\maketitle

\section{Introduction}

Single molecule biochemical experiments provide highly detailed
knowledge about the mean time between successive reaction events and
hence about the reaction rates. Additionally, they deliver
qualitatively new information, inaccessible to bulk experiments, by
measuring other reactions statistics, such as variances and
autocorrelations of successive reaction times
\cite{english-00,orrit,gopich-03,english-06,xue-06,chaudhury-07}. In
their turn, these quantities relate to structural properties of the
reaction networks, uncovering such phenomena as internal enzyme states
or multi-step nature of seemingly simple reactions, and hence starting
a new chapter in the studies of the complex biochemistry that
underlies cellular regulation, signaling, and metabolism.

However, the bridge between the experimental data and the network
properties is not trivial. Since the class of exactly solvable
biologically relevant models is limited
\cite{hornos-05,darvey-66,Jayaprakash-07}, exact analytical
calculations of statistical properties of reactions are impossible
even for some of the simplest networks.  Similarly, the variational
approach \cite{sasai-03,lan-06} and other analytical approximations
are of little help when the experimentally observed quantities depend
on features that are difficult to approximate, such as the tails of
the reaction events distributions. Therefore, computer simulations are
often the method of choice to explore an agreement between a presumed
model and the observed experimental data.

Unfortunately, even the simplest biochemical simulations often face
serious problems, both conceptual and practical. First, the networks
usually involve {\em combinatorially many} molecular species and
elementary reaction processes: for example, a single molecular
receptor with $n$ modification sites can exist in $2^n$ states, and an
even larger number of reactions connecting them
\cite{hlavacek}. Second, while it is widely known that {\em some}
molecules occur in the cell at very low copy numbers (e.g., a single
copy of the DNA), which give rise to relatively large stochastic
fluctuations\cite{low-copy1,low-copy2,low-copy3,low-copy4,low-copy5,low-copy6,low-copy7},
it is less appreciated that the combinatorial complexity makes this
true for {\em almost all} molecular species. Indeed, complex systems
with a large number of molecules, like in eukaryotic cells, may have
small abundances of typical microscopic species if the number of the
species is combinatorially large. Third, and perhaps the most profound
difficulty of the ``understanding-through-simulations'' approach, is
that only very few of the kinetic parameters underlying the
combinatorially complex, stochastic biochemical networks are
experimentally observed or even observable. For example, the average
rate of phosphorylation of a receptor on a particular residue can be
measured, but it is hopeless to try to determine the rate for each of
the individual microscopic states of the molecule determined by its
modification on each of the other available sites.

While some day computers may be able to tackle the formidable problem
of modeling astronomically large biochemical networks as a series of
random discrete molecular reaction events (which will properly account
for stochastic copy number fluctuations), and then performing sweeps
through parameter spaces in search of an agreement with experiments,
such powerful computers are still far away. More importantly, even if
this computational ability were available, it would not help in
building a comprehensible, tractable interpretation of the modeled
biological processes and in identifying connections between
microscopic features and macroscopic parameters of the networks.

\begin{figure}[t]
\centerline{\includegraphics[width=14 cm]{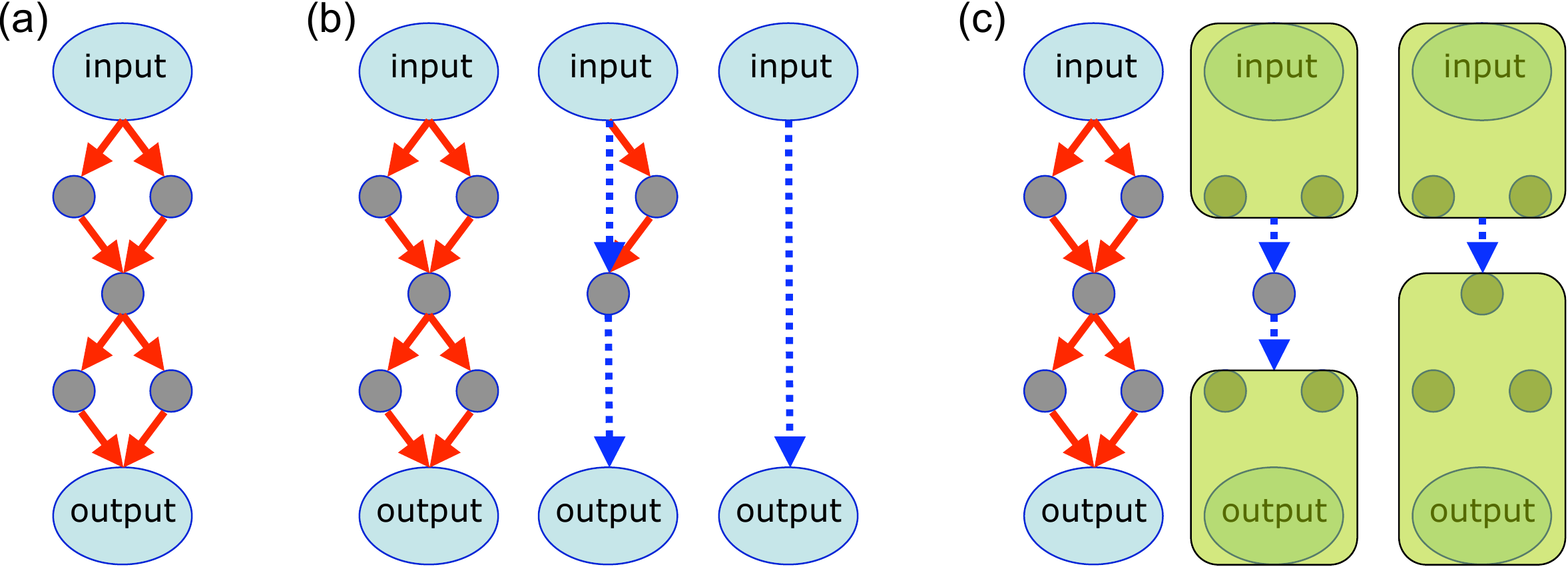}}
\caption{\label{cascade} (a) A complex network of elementary reactions
  connecting the {\em input} and the {\em output} nodes. Note that the
  choice of these nodes usually distinguishes different
  coarse-graining schemes, and it is rather arbitrary. In our work,
  the choice is determined by the adiabatic time scale separation, as
  described in {\em Methods}. In principle, such networks can be
  coarse-grained by multiple methods. In (b) we illustrate the
  decimation procedure, where intermediate nodes with fast dynamics
  get eliminated successively, resulting in complex reactions
  connecting their immediate neighbors; the statistics of these
  complex reactions are determined by the cumulant generating
  functions (CGF) ${\mathcal S}(\chi)$, cf.~{\em Results}. Other
  coarse-graining schemes are possible. For example, (c) nodes can be
  merged in hyper-nodes, again connected to each other by complex
  reactions. Combinations of the strategies are also allowed. Panels
  (b) and (c) resemble the decimation and the blocking procedures in
  statistical physics \cite{kadanoff}, and, not coincidentally,
  statistical physics is the field where coarse-graining has had the
  biggest impact and is the most developed \cite{ma}. Both the
  decimation and the blocking are in the spirit of the real-space
  renormalization group on an irregular lattice, and one can also
  think of momentum-space-like approaches as a complement \cite{ma}. }
\end{figure}

Clearly, such an interpretation can be aided by simplifying the
networks through coarse-graining, that is, by merging or eliminating
certain nodes and/or reaction processes. Ideally, as in
Fig.~\ref{cascade}, one wants to substitute a whole network of
elementary (that is, single-step, Poisson-distributed) biochemical
reactions with a few complex reaction links connecting the species
that survive the coarse-graining in a way that retains predictability
of the system.  Not incidentally, this would also help with each of
the three major roadblocks mentioned above, by reducing the number of
interacting elements, increasing the number of molecules in
agglomerated hyperspecies, and combining multiple features into a much
smaller number of effective, mesoscopic kinetic parameters.

The importance of coarse-graining in biochemistry has been understood
since 1913 \cite{MM}, when the first deterministic coarse-graining
mechanism, now known as the Michaelis-Menten (MM) enzyme, was
proposed for the following kinetic scheme:
\begin{equation}
E+S \xrightleftharpoons[k_{-1}{[C]}]{k_{1}[E][S]}
C\xrightarrow{k_2[C]} E+P.
\label{MM-1}
\end{equation}
Here $k_1$, $k_2$, and $k_{-1}$ are kinetic rates, $S$, $P$, $E$, and
$C$ denote the substrate, the product, the enzyme, and the
enzyme-substrate complex molecules, respectively, and $[\dots]$
represent the abundances.  The enzyme catalyzes the $S\to P$
transformation by merging with $S$ to create an unstable complex $C$,
which then dissociates either back into $E+S$ or forward into $E+P$,
leaving $E$ unmodified.  If $[S]\gg[E]$, then the enzyme cycles many
times before $[S]$ and $[P]$ change appreciably. This allows to
simplify the enzyme-mediated dynamics by assuming that the enzymes
equilibrate quickly at the current substrate concentration, resulting
in a coarse-grained, complex reaction with decimated enzyme species:
\begin{equation}
S \xrightarrow{v} P,\quad
v=\frac{k_2[S][E]}{[S]+(k_2+k_{-1})/k_1}.
\label{MM-rate}
\end{equation}

However, this simple reduction is insufficient when stochastic effects
are important: each complex MM reaction consists of multiple
elementary steps, thus the statistics of the number of MM reactions
per unit time, in general, is non-Poissonian. While some relatively
successful attempts have been undertaken to extend simple
deterministic coarse-graining to the stochastic domain
\cite{arkin-1,arkin-2,arkin-3,gopich-06,sinitsyn-07epl}, a general set
of tools for coarse-graining large biochemical networks has not been
found yet.

In this article, we propose a method for a systematic rigorous
coarse-graining of stochastic biochemical networks, which can be
viewed as a step towards creation of comprehensive coarse-grained
analysis tools. We start by noting that, in addition to the conceptual
problems mentioned above, a technical difficulty stands in the way of
stochastic simulations methods in systems biology: molecular species
and reactions have very different dynamical time scales, which makes
biochemical networks stiff and difficult to simulate. Here we propose
to use this property of separation of time scales to our advantage.

The idea is not new, and multiple related approaches have been
proposed in the literature, differing from each other mainly in the
definition of fast and slow variables.  A common practice is to use
{\em reaction rates} to identify fast and slow reactions
\cite{arkin-1,arkin-2,arkin-3}.  However, if two species of very
different typical abundances are coupled by one reaction, then a
relatively small change in the concentration of the high abundance
species can have a dramatic effect on that of the low abundance one.
This notion of {\em species}-based rather than {\em reaction}-based
adiabaticity has been used in the original MM derivation, and it is
also at the heart of our arguments.

Our method builds upon the stochastic path integral technique from
mesoscopic physics
\cite{pilgram-03,pilgram-04,elgart-04,sinitsyn-07prl}, providing three
major improvements that make the approach more applicable to
biological networks.  First, we extend the method, initially developed
for large copy number species, to deal with simple discrete degrees of
freedom, such as a single MM enzyme or a single gene.  Second, we
explain how to apply the technique to a network of multiple reactions,
thereby reducing the entire network to a single complex reaction
step. Finally, we show how the procedure can be turned into an
efficient algorithm for simulations of coarse-grained networks, while
preserving important statistical characteristics of the original
dynamics. The algorithm is akin to the Langevin \cite{zinn-justin} or
$\tau$-leaping \cite{tau-leapin1,tau-leapin2} schemes, widely used in
biochemical simulations, but it allows to simulate an entire complex
reaction in a single step. We believe that this development of a fast,
yet precise simulation algorithm is the most important practical
contribution of this manuscript.

For pedagogical reasons, we develop the method using a model system
that is simple enough for a detailed analysis, yet is complex enough
to support our goals. A generalization to more complex systems is
suggested in the {\em Discussion}.

\subsection{The model}
\begin{figure}[t]
\centerline{\includegraphics[width=8 cm]{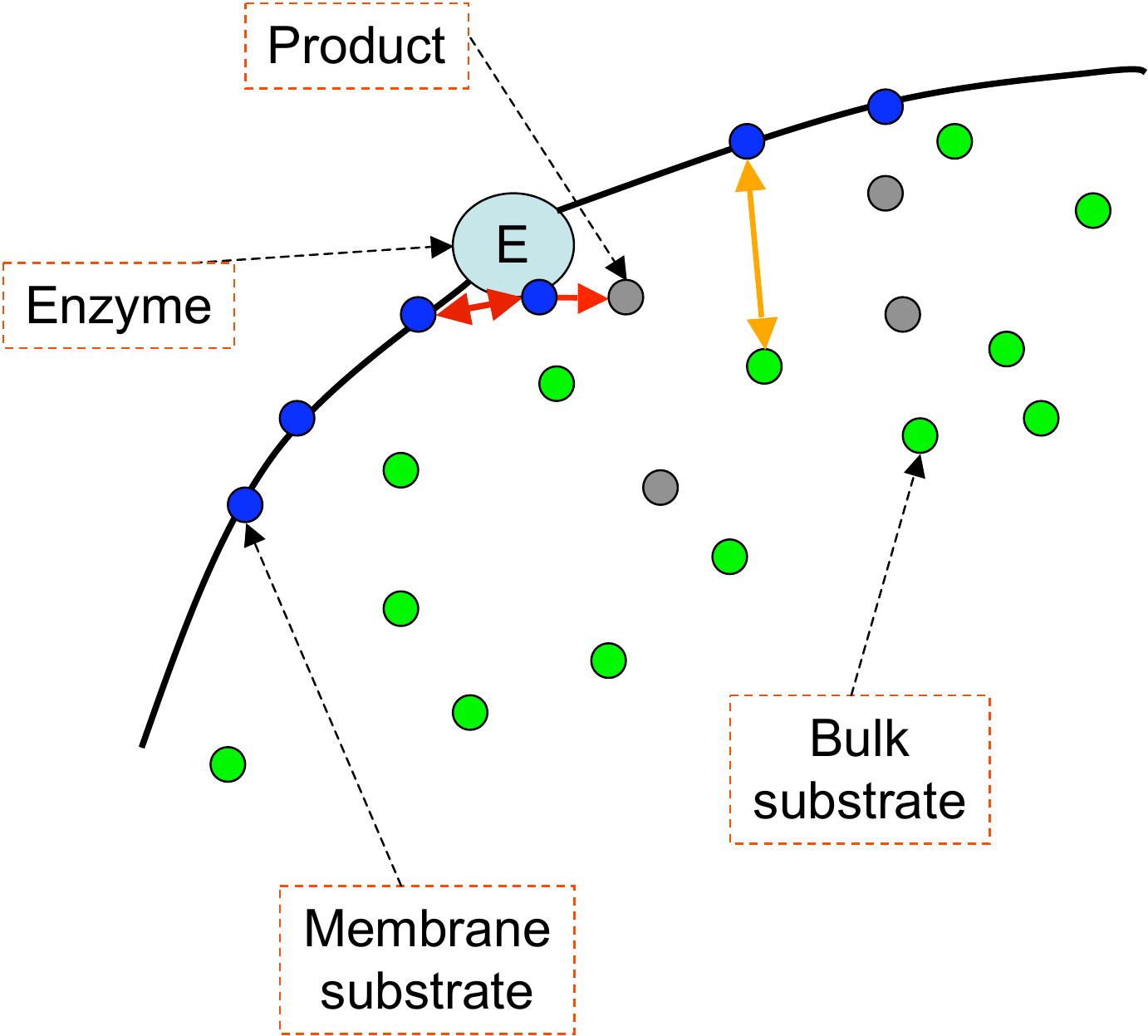}}
\caption{\label{system} A single enzyme on a membrane, interacting
  with substrate molecules. Green, blue, and grey circles are bulk
  substrate, membrane substrate, and product molecules,
  respectively. Arrows represent possible reactions: (1,2) adsorption
  and dissociation of $S$ by/from the membrane (orange); (3)
  multi-step MM conversion $S\to P$ (red).}
\end{figure}

Consider the system in Fig.~\ref{system}: an enzyme is attached to a
membrane in a cell.  $S_{\rm B}$ substrate molecules are distributed
over the bulk cell volume. Each molecule can either be adsorbed by the
membrane, forming the species $S_{\rm M}$, or dissociate from it.
Enzyme-substrate interactions are only possible in the adsorbed
state. One can easily recognize this as an extremely simplified model
of receptor mediated signaling, such as in vision
\cite{GFP,detwiler-00}, or immune signaling.

As usual, the enzyme-substrate complex $C$ can split either into $E+S$
or into $E+P$.  Let's suppose that the latter reaction is observable;
for example, a GFP-tagged enzyme sparks each time a product molecule
is created \cite{english-06}. We further suppose the reaction $C\to
E+P$ is not reversible (that is, the product leaves the membrane or
the reaction requires energy and is far from equilibrium).

The full set of elementary reactions is 
\begin{enumerate}
\item adsorption of the bulk substrate onto the membrane, $S_{\rm B}
\rightarrow S_{\rm M}$, with rate $q_0S_{\rm B}$;
\item reemission of the substrate back into the bulk, $S_{\rm M}
\rightarrow S_{\rm B}$, with rate $qS_{\rm M}$;
\item Michaelis-Menten conversion of $S_{\rm M}$ into $P$, consisting of
\begin{enumerate}
\item substrate-enzyme complex formation, $S_{M}+E \rightarrow C$,
  with rate $k_1S_{\rm M}$;
\item complex backward decay, $C \rightarrow S_{\rm M}+E$, with rate
  $k_{-1}$;
\item product emission $C \rightarrow E+P$, with rate $k_{2}$.
\end{enumerate}
\end{enumerate}
Note that here and in the rest of the article, we drop the $[\dots]$
notation for denoting abundances and don't make a distinction between
a species name and the number of its molecules.

In this setup, only emission of the product is directly
observable. Our goal is to coarse-grain the above system of five reaction
processes into a single complex reaction $S_{\rm B}\to P$, as in
Fig.~\ref{reduction}(c). That is, we want to eliminate all
intermediate species and reaction processes, while preserving their
effects on the statistical properties of the complex reaction $S_{\rm
  B}\to P$ on time scales appropriate for its dynamics.


This set of reactions has another interpretation. Consider an MM
enzyme in the bulk together with the substrate.  When the substrate
concentration is small, only few of its molecules are sufficiently
close to the enzyme to interact with it. In this context, one can
approximate the full reaction-diffusion setup by a system having an
inner (reactive) and an outer (non-reactive) regions surrounding the
enzyme. Diffusion takes substrate molecules between the regions with
(almost) Poisson statistics of transitions, with the rate parameters
depending on the volume of the regions and the diffusion coefficient.
The particles in the inner region can interact with the MM enzyme,
completing the mapping between the reaction-diffusion system and the
multi-state well-mixed kinetic process described above.

\section{Results}

There are three distinct effective time scales in the system in
Fig.~\ref{system}. One is the time scale $\tau_{\rm B}$ of the
variation of the bulk substrate abundance.  We assume that $S_{\rm B}$
is much larger than $S_{\rm M}$. Therefore, this time scale is the
slowest, and we will be interested in studying the response of the
system to the bulk substrate abundance $S_{\rm B}$ on this scale. A
faster time scale is given by the dynamics of the molecules on the
membrane, $\tau_{\rm M}$. Finally, at the other extreme, the fastest
time scale, $\tau_{\rm E}$, is set by single reaction events, that is,
the characteristic time between two successive product releases by the
enzyme. Overall, $\tau_{\rm E}\ll\tau_{\rm M}\ll\tau_{\rm B}$.

We emphasize again that all species in the problem are connected by
reactions that happen with approximately the same rates, and the
separation of the time scales is a direct result of the particle
abundances, rather than the conversion speeds: it takes only a few
reaction events to change a low-abundance species drastically, and a
lot longer to do the same to a high-abundance one. This is the main
reason why we believe that this illustrative model will shed light on
coarse-graining of a wide class of networks.

\begin{figure}[t]
\centerline{\includegraphics[width=10 cm]{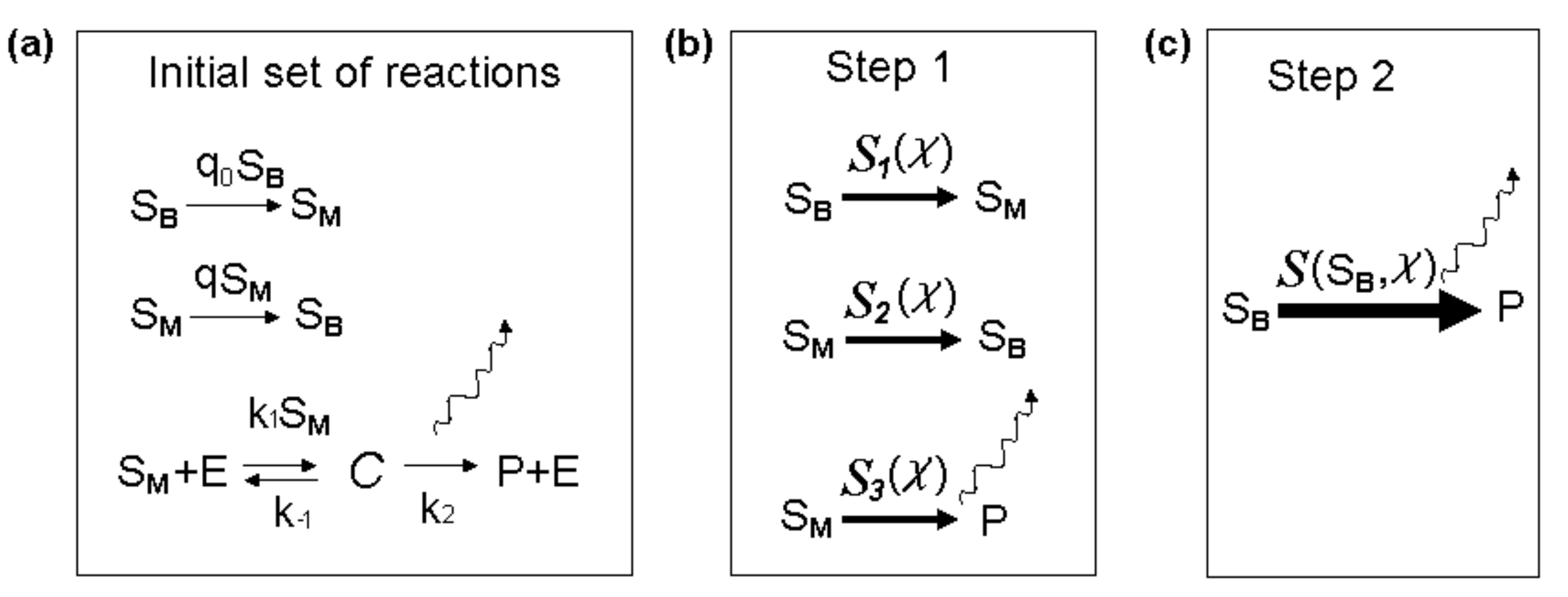}}
\caption{\label{reduction} Coarse-graining of the model. In (a) we
  show the original set of reactions describing the system in
  Fig.~\ref{system}. Panel (b) represents the set of reactions after
  the first coarse-graining step. Note that here the three stages of
  the MM enzyme have been replaced by a single complex
  reaction. Further, all the remaining (simple or complex) reactions
  are now represented by their slowly varying CGFs. Panel (c) shows
  the final reaction that describes the system at time scales much
  larger than the characteristic time of evolution of $S_{\rm M}$, the
  number of the substrate molecules on the membrane. The wavy line
  corresponds to a spark of the tracer molecule \cite{english-06},
  which counts the number of $S_{\rm B}\to P$ transformations.}
\end{figure}

The hierarchy of times allows us to coarse-grain the system in two
steps, as in Fig.~\ref{reduction}. First, we remove the variable with
the fastest dynamics, that is, the binary occupancy substrate-enzyme
complex $C$. This replaces the three reactions of the MM mechanism
with a single reaction $S_{\rm M}\to P$
(Fig.~\ref{reduction}(b)). Additionally, we represent the other
reactions in the system in a form more suitable for the subsequent
developments. In the second step, we eliminate the membrane-bound
substrate variable, which evolves on the scale $\tau_{\rm M}$. This
results in the characterization of the average flux and its
fluctuations for $S_{\rm B}\to P$ transformation, treating $S_{\rm B}$
as a time-dependent input parameter, cf.~Fig.~\ref{reduction}(c).


\subsection{Preliminaries}
Let $\delta Q_{\mu}$ stand for the number of reaction events for the
reaction type $\mu$ (in our example, $\mu=1,2,3$ corresponds to
adsorption, detachment, and the whole MM reaction, respectively). Then
$P(\delta Q_{\mu}|T)$ is the probability distribution of the number of
events of type $\mu$ during a temporal window of length $T$.  
Instead of considering these distributions directly, our derivation for
the coarse-graining relies the corresponding
moment generation functions (MGFs)\footnote{More precisely, ${\mathcal
    Z}$ is the {\em characteristic} function, and the usual definition
  of the MGF is without $i$ in the exponent. Same is true for our
  somewhat incorrectly named CGF, ${\mathcal S}$. We choose this
  unconventional nomenclature to emphasize that our main use of the
  functions is for calculations of moments and cumulants,
  respectively.}:
\begin{equation}
 {\mathcal Z}_{\mu}(\chi,T)=e^{{\mathcal S}_{\mu}(\chi,T)}=
 \sum_{\delta Q_{\mu}=0}^{\infty} P(\delta Q_{\mu}|T)e^{i\delta Q_{\mu}\chi}.
\label{pgf111}
\end{equation}
where ${\mathcal S}_{\mu}(\chi)$ is the cumulant generating function (CGF). The
moments and the (complex) cumulants of the distribution $P(\delta Q_{\mu}|T)$ can be
calculated by differentiating the MGF and the CGF, respectively
\begin{eqnarray}
m_{\mu,a}&=&(-i)^a\left.\frac{\partial^a}{\partial \chi^a}\right|_{\chi=0}
{\mathcal Z}_{\mu}(\chi,T)\\
c_{\mu,a}&=&(-i)^a\left.\frac{\partial^a}{\partial \chi^a}\right|_{\chi=0}
{\mathcal S}_{\mu}(\chi,T),\label{cumulants}
\end{eqnarray}
where $a$ stands for the order of the moment (cumulant). In
particular, the average flux for the reaction $\mu$ is
$c_{\mu,1}=m_{\mu,1}$, and the corresponding variance is
$\sigma^2_{\mu}=c_{\mu,2}=m_{\mu,2}-m_{\mu,1}^2$.

\subsection{Step 1: The generating function representation}

This step can be viewed as a generalization of the standard
$\tau$-leaping approximation \cite{tau-leapin1,tau-leapin2}, which
prescribes to simulate elementary, exponentially distributed reaction
events, for example attachment/detachment reactions in
Fig.~\ref{reduction}(a), by choosing a time step such that the number
of the reactions in it is much larger than unity, yet the reaction
rates (determined by the dynamics of the slower variables in the
problem) can be considered stationary. In a $\tau$-leaping scheme, one
then approximates the distribution of the number of reaction events by
a Poisson distribution.

Unfortunately, not all biochemical processes can be treated in such a
simple manner. For example, due to the single-copy nature of the MM
enzyme in our system, Fig.~\ref{reduction}(a), the instantaneous rate
of the product creation is a fast varying function of time, switching
between zero and $k_2$ every time binding/unbinding events
happen. Therefore, one cannot use $\tau$-leaping or Langevin schemes
\cite{zinn-justin,tau-leapin1,tau-leapin2}, or treat the product
creation as a homogeneous Poisson process. We would like to avoid being
forced into the Gillespie \cite{gillespie} or the StochSim
\cite{StochSim} analysis schemes.  Since either of these schemes
is based on Monte-Carlo
simulations of every individual reaction event, the estimation
of parameters of interest may become excessively
slow in large systems. 

As an alternative,  we will identify
good approximations for the distribution of the number of reactions in 
a fixed time interval, which is no longer a Poisson distribution, 
by characterizing its CGF (see {\em Methods: Moment
  generating functions for elementary chemical reactions}).
To this end, we propose to eliminate the binary
substrate-enzyme complex variable and reduce the MM reaction triplet
to a single reaction, whose dynamics can be considered stationary over
times much longer than a single reaction event.  This completes Step 1 of the coarse-graining in which each reaction,
or a small complex of reactions, is subsumed by a CGF ${\mathcal S}_{\mu}$
of the distribution of the number of events,
which can be considered stationary for extended times.  Importantly,
in this Step, we removed the only species in the problem that exists,
at most, in a single copy and hence is stiff.  This dramatically
simplifies simulations and analysis of the system.

The details of this
are given in {\em Methods: Coarse-graining the Michaelis-Menten
  reaction}. In particular, in Eq.~(\ref{s3}) we derive ${\mathcal
  S}_3$, the CGF for the entire complex Michaelis-Menten reaction,
eliminating the intermediate substrate-enzyme complex $C$. The
expression is valid over times much larger than $\tau_{\rm E}$, but
smaller than $\tau_{\rm M}$, so that many enzyme turnovers happen, but
the effect on the abundance of the membrane-bound substrate is still
relatively small.

In the MM mechanism, the backward reaction is often a
simple dissociation, whereas the forward one requires crossing an
energy barrier and is exponentially suppressed. As a result, one often
has $k_{-1}\gg k_2$, which can make the MM reaction doubly stiff,
requiring multiple binding events (and with them the instantaneous
rate changes) for each released product. Therefore, replacing the
entire MM mechanism with a single complex reaction step has a dramatic
effect on analysis of the reaction, and specifically on the simulation
efficiency, which can now be performed using the Langevin-like
quasi-stationary approximation.

\begin{figure}[t]
\centerline{\includegraphics[width=12 cm]{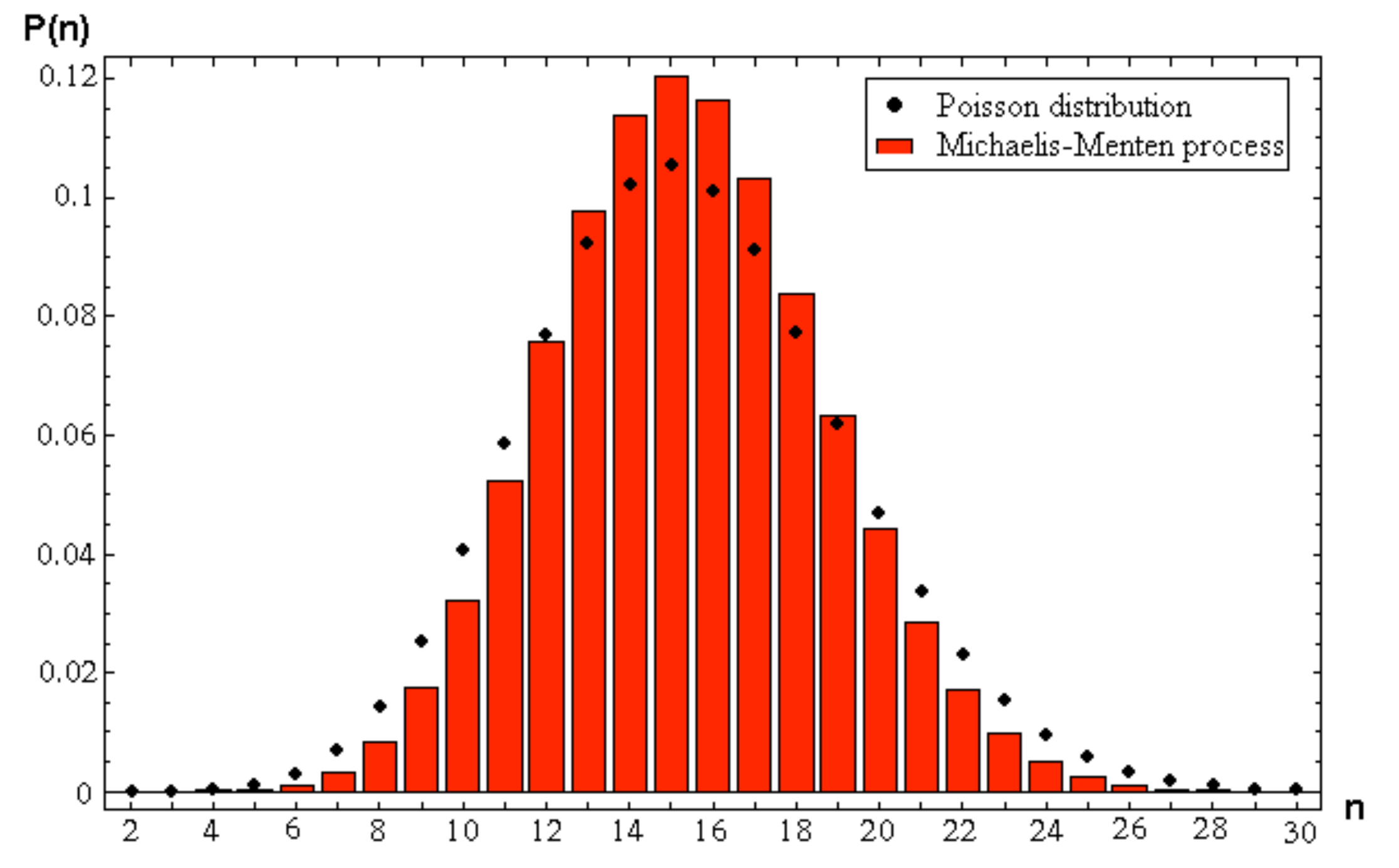}}
\caption{\label{MMdistr} Distribution of the number of
  Michaelis-Menten reactions over a time $\delta t=35$ with $S_{\rm
    M}=140$, $k_1=0.01$, $k_{-1}=1$, and $k_2=1$ vs.\ the Poisson
  distribution with the same average number of reactions. The
  distribution for the MM process is obtained using the Gram-Charlier
  expansion with the four known cumulants, see {\em Methods}.}
\end{figure}

To illustrate this, using Eq.~(\ref{s3}), we write the first few
cumulants of the number of MM product releases per time $\delta t$:
\begin{eqnarray}
 \mu_3&=& \frac{k_1k_2S_{\rm M}}{K}\,\delta t,\quad 
 K=k_1S_{\rm M}+k_2+k_{-1},  \label{MMmean}\\
 \sigma^2_3&=&\mu_3 F, \quad F=\left(1- \frac{2k_1k_2S_{\rm M}}{K^2} \right) , 
 \label{MMsigma}\\
 c_{3,3}&=&\mu_3 \left( 1-\frac{6Q(K-2Q)}{K^2} \right) , \quad Q=\mu_3/\delta t, \label{MM3}\\
 c_{3,4}&=&\mu_3 \left( 1- 
 \frac{2Q(7K^2-36KQ +60Q^2 }{K^3} \right).\label{MM4}
\end{eqnarray}
The coefficient $F$ in the expression for $\sigma^2_3$ is called the
Fano factor (see below).  To the extent that $\sigma^2_3\neq\mu_3$,
this complex reaction is non-Poisson, as illustrated in
Fig.~\ref{MMdistr}.

Knowing cumulants of the reaction events distribution allows for a
simple numerical simulation procedure
\begin{eqnarray}
\delta Q_3(t)&=&  \eta_3(t,\delta t),\label{mm_sim1}\\
S_{\rm M}(t+\delta t)&=&S_{\rm M}(t)-\delta Q_3(t)+J(t)\delta t,\label{mm_sim2}\\
P(t+\delta t)&=&P(t)+\delta Q_3(t)\label{mm_sim3},
\end{eqnarray}
where $\eta_3(t)$ is a random variable with the cumulants given by
Eqs.~(\ref{MMmean}-\ref{MM4}), and $J(t)$ represents currents
exogenous to the MM reaction, such as changes in $S_{\rm M}$ due to
membrane binding/unbinding.  Notice that, in this step, we are now
treating the single-particle-mediated reaction in a quasi-stationary,
$\tau$-leaping or Langevin-like way by drawing a (random) number of
complex reaction events over a time $\delta t$ directly, assuming that
all parameters defining the reaction are constants over this time.
The price for the coarse graining is that instead of characterizing
any reaction by a single rate that defines a Poisson distribution of
reaction events, one is forced to use a distribution with the
prescribed sequence of moments for the Monte-Carlo simulations.

In principle, generation of such random variables is a difficult and
an ill-posed task since the moments do not define the distribution
uniquely, and two distributions with matched moments can be arbitrarily
different from each other.  Additionally, once we allow for a nonzero
third or fourth cumulant, the remaining higher order cumulants cannot
be zero anymore\cite{lax}, and the generated random variable will
depend on the assumptions made about them. Fortunately, in our case,
the situation is simplified because all $c_{3,k}\sim \delta t$. Thus
the $k$'th cumulant will have a progressively smaller effect,
$\sim(\delta t)^{1/k}$, on a number drawn from the distribution, and
our random variables are almost Gaussian. Then the Gram-Charlier series
expansion \cite{edgeworth} aided either by  the importance or rejection sampling
\cite{imp-sampl} reduces the simulation scheme,
Eqs.~(\ref{mm_sim1}-\ref{mm_sim3}), to a simple Langevin simulation
with a Gaussian noise and a small penalty, as described in
{\em Methods: Simulations with near-Gaussian distributions}; see also
Fig.~\ref{comparison} in {\em Methods} for illustration of the
precision provided by these tools.

\subsection{Step 2: Coarse-graining membrane reactions}

For Step 2 of our coarse-graining approach, we are given the CGFs
$S_{\mu}$, $\mu=1,2,3$, of the slowly varying reactions.  Using the
stochastic path integral technique, we then express the CGF of the
entire coarse-grained reaction $S_{\rm B}\to P$ in
Fig.~\ref{reduction}(c) in terms of the component CGFs, and then
simplify the expression to account for the time scale separation
between the dynamics of $S_{\rm B}$ and $S_{\rm M}$. This is presented
in detail in {\em Methods: Coarse-graining all membrane reactions},
cf.~Eq.~(\ref{fcs7}). This formally completes the coarse-graining. That
is, we find the CGF of the $S_{\rm B}\to P$ particle flux for times
$T\lesssim \tau_{\rm B}$, much longer than $\tau_{\rm E}$ and
$\tau_{\rm M}$, the other time scales in the problem.

The resulting CGF depends on microscopic reaction rates, which can
depend on slow parameters, such as $S_{\rm B}$. The full expression
for CGF is cumbersome and non-illuminating. Fortunately, we only want
to calculate the first few cumulants of the reaction events
distribution, and these are obtained by differentiation the CGF as in
Eq.~(\ref{cumulants}).  This gives
\begin{eqnarray}
 c_1&=& T\frac{1}{2k_1} \bigg[ k_1(k_0+k_2)+q(k_2+k_{-1})
 \nonumber\\
 &&-
   \sqrt{k_1^2(k_0 - k_2)^2 + 2k_1 q (k_0 +k_2)(k_2
     +k_{-1})+q^2 (k_2+k_{-1})^2} \bigg],
\label{jp4} 
\label{c1_formula}\\
c_2&=&Fc_1,\label{c2_formula}\\ \quad 
&&F=1-\frac{q(2k_1 k_0 k_2 +k_1 (k_0+k_2)k_{-1}+q
  k_{-1}(k_2+k_{-1}))}
{k_1^2 (k_0-k_2)^2 + 2 k_1 q (k_0+k_2)(k_2+k_{-1})+q^2(k_2+k_{-1})^2} + \nonumber\\
&&\quad\quad\quad\quad+\frac{qk_{-1}}{\sqrt{k_1^2(k_0 - k_2)^2 + 2k_1 q (k_0 +k_2)(k_2 +k_{-1})+q^2 (k_2+k_{-1})^2}}.
\label{fano_f}
\\
c_3&=&-T\frac{\kappa}{\rho(-\kappa k_1+\rho^2)^5} \left\{\kappa^5k_1^{5}-\rho^{10}+
    \kappa \rho^7\left[5k_1^2k_2+q\left(11k_1+6q\right)s\right]\right.\nonumber\\
&&-\kappa^2k_0^2k_1^4\rho^2\left[5k_1^2k_2^2+6k_2\left(k_1-2q\right)qs+24q^2s^2 \right]\nonumber\\
&&+2\kappa^2k_0k_1^2\rho^3\left[5k_1^3k_2^2+k_2q\left(14k_1^2-9k_1q-6q^2\right)s+6q^2\left(5k_1+3q\right)s^2\right]\nonumber\\
&&\left.-2\kappa k_0k_1\rho^4\left[5k_1^4k_2^3+19k_1^3k_2^2qs+
9k_1^2k_2q^2s^2+ 6k_2q^4s^2+3k_1q^3s\left(-2k_2^2+
8k_2s+s^2\right)\right]\right\},
\label{c3_formula}
\end{eqnarray}
where $s=k_1 \la S_{\rm M} \ra+k_2+k_{-1}$, $\la S_{\rm M}
\ra=\frac{1}{2k_1 q}\Big\{ k_0 k_1 -k_1 k_2 - k_2 q -k_{-1} q +
\big[4k_1 k_0 q (k_2 + k_{-1})+(k_1 k_2 -k_1 k_0 +k_2 q +k_{-1}q
)^2\big]^{1/2}\Big\}$ is the average number of membrane-bound
substrates, $k_0=q_0S_{\rm B}$, $\kappa=k_0k_1k_2$, $\rho=k_1k_2+qs$,
and, finally, $T$ is the time step over which $S_{\rm B}$ changes by a
relatively small amount, but many membrane reactions
happen.

By analogy to the MM reaction, Eqs.~(\ref{mm_sim1}-\ref{mm_sim3}),
the results from Step 2 allow for simulations of the whole reaction scheme in one
Langevin-like step:
\begin{eqnarray}
 \delta Q (t)&=& \eta(t,T),\\
 S_{\rm B}(t+T)&=&S_{\rm B}(t)-\delta Q(t) + J(t)T,\\
 P(t+T)&=&P(t)+\delta Q(t),
\end{eqnarray}
where $\eta$ is a random variable with the cumulants as in
Eqs.~(\ref{c1_formula}-\ref{c3_formula}), to be generated as in {\em
 Methods: Simulations with near-Gaussian distributions}, and $J(t)$
is an external current, such as production or decay of the bulk
substrate in other cellular processes.

\subsection{Fano factor in a single molecule experiment}
In analyses of single molecule experiments, one often calculates the
ratio of the variance of the reaction events distribution to its mean,
called the Fano factor \cite{english-06,fano-ref1}:
\begin{equation}
F=\frac{c_2}{c_1}.
\end{equation}
The Fano factor is zero for deterministic systems and one for a totally
random process described by a Poisson number of
reactions.  As a result, the Fano factor provides a natural quantification of the
importance of the stochastic effects in the studied process. In vivo,
it can be measured, for example, by tagging the enzyme with a
fluorescent label that emits light every time a product molecule is
released \cite{english-06}.

\begin{figure}[t]
\centerline{\includegraphics[width=10 cm]{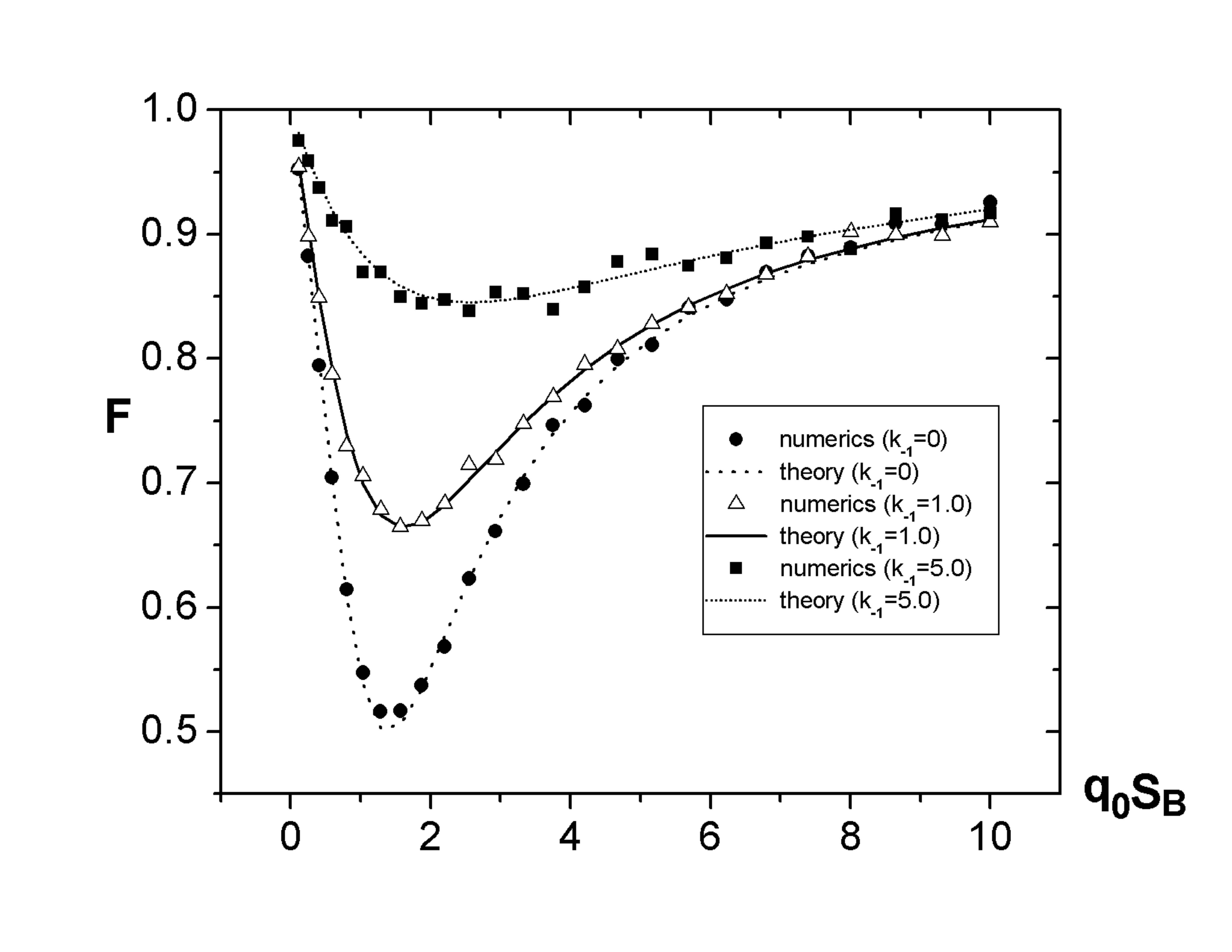}}
\caption{\label{Fano factor} Comparison of the analytically calculated
  Fano factor for the $S_{\rm B}\to P$ reaction, Eq.~(\ref{fano_f}), to
  direct Monte Carlo simulations with the Gillespie algorithm
  \cite{gillespie}. Here we use $q=0.02$, $k_1=0.05$, $k_2=1$,
  evolution time $T=10000$ (in arbitrary units).  Each numerical data
  point was obtained by averaging 10000 simulation runs.}
\end{figure}

Traditionally, to compare experimental data to a mathematical model,
one would simulate the model using the Gillespie kinetic Monte Carlo
algorithm \cite{gillespie}, which is a slow and laborious process that takes
a long time to converge to the necessary accuracy (see
below).   In contrast, our coarse-graining approximations yields 
an analytic expression for the Fano factor of the $S_{\rm B}\to P$
transformation via Eq.~(\ref{fano_f}).  This illustrates a first 
practical utility of our coarse-graining approach.
In Fig.~\ref{Fano factor}, we compare this analytical
expression, derived under the aforementioned quasi-stationary
assumption, with stochastic simulations for the full set of elementary
reactions in Fig.~\ref{reduction}(a). The results are in an excellent
agreement, illustrating the power of the analytical approach.

Note that the Fano factor is generally different from unity,
indicating a non-Poissonian behavior of the complex reaction. The
backwards decay of $C$, parameterized by $k_{-1}$, adds extra
randomization and thus larger values of $k_{-1}$ increases the
Fano factor $F$. At
another extreme, when $k_{-1}=0$, the Fano factor may be as small as
1/2, indicating that then the entire $S_{\rm B}\to P$ chain can be
described by the sum of two Poisson events with similar rates, which halves
the Fano factor.  Finally, when $q=0$, i.e., the substrates are
removed from the membrane only via conversion to products, the
Fano factor $F$ is one. This is because here the only stochasticity in the
problem arises from Poisson membrane binding: on long time-scales, all
bound substrates will eventually get converted to the outgoing flux.

\subsection{Computational complexity of coarse-grained simulations}

As we alluded to before, in addition to analytical results, such as
the expression for the Fano factor, we expect the coarse-graining
approach to be particularly useful for stochastic
simulations in systems biology. This is due to an essential speedup
provided by the method over traditional simulation techniques, such
as, in particular, the Gillespie algorithm \cite{gillespie}, to which
most other approaches are compared too. 

Indeed, for the model analyzed in this work, the computational
complexity of a single Gillespie simulation run is
$O\left(M\frac{T}{\tau_{\rm E}}\right)$, where $M=5$ is the number of
reactions in the system, and $T$ is the duration of the simulated
dynamics. In contrast, the complexity of the coarse-grained approach
scales as $O\left(M^0\left(\frac{T}{\tau_{\rm E}}\right)^0\right)$
since we have removed the internal species and simulate the dynamics
in steps of $\sim T$, instead of steps $\sim\tau_E$. However, since the
coarse-grained approach requires generation of complicated random
numbers, the actual reduction in the complexity is unclear. More
importantly, the Gillespie algorithm is (statistically) exact, while
our analysis relies on quasi-stationary assumptions. Therefore, to
gauge the practical utility of our approach in reducing the simulation
time while retaining a high accuracy, we benchmark it against the
Gillespie algorithm. We do this for a single MM enzyme first, and then
progress to the full five reaction model of the enzyme on a
membrane. Details of the computer system used for the benchmarking can
be found in {\em Methods: Simulations details}.

{\em The Michaelis-Menten model:} We consider a MM enzyme with
$S_{\rm M}=140={\rm const}$, $k_1=0.01$, $k_{-1}=2.0$, $k_2=1.0$. 
We analyze the number of product molecules produced by this
enzyme over time $\delta t=35$, with the enzyme initially in the
(stochastic) steady state. To strain both methods, we require a very
high simulation accuracy, namely convergence of the fourth moment of
the product flux distribution to two significant digits. For both
methods, this means over 10 millions realizations of the same
evolution.

\begin{table}[t]
\begin{tabular}{|c|cc|c|}
 \hline
 Cumulants &  Gillespie    &   Coarse-grained  & Analytical prediction  \\
 \hline
 $c_1$   & 11.24 $\pm$ 0.01 & 11.14 $\pm$ 0.01  & 11.14  \\
 $c_2/c_1$   & 0.843 $\pm$ 0.001 & 0.855 $\pm$ 0.001 & 0.855 \\
 $c_3/c_1$  & 0.613 $\pm$ 0.004 & 0.628 $\pm$ 0.004 & 0.628 \\
 $c_4/c_1$   & 0.32 $\pm$ 0.02 & 0.32 $\pm$ 0.02 & 0.319   \\
 time   & 8 min 45 s & 12 s & N/A   \\
 \hline
\end{tabular}
\caption{\label{table_MM}Comparison of the Gillespie and the
  coarse-grained simulation
  algorithms. The numbers are reported for 12 million realizations of
  the same evolution for each of the methods. To highlight deviations
  from the Poisson and the Gaussian
  statistics, we provide ratios of the higher order cumulants to the
  mean of the product flux distribution. In the last column, we report
  analytical predictions, Eqs.~(\ref{MMmean}-\ref{MM4}), obtained from
  the quasi-steady state approximation to the CGF.}
\end{table}

In Tbl.~\ref{table_MM} we report the results of our simulations.  We
see that the analytical coarse-grained results differ from the exact
Gillespie simulations by, at most, two per cent, which is an expected
deviation given the quality of the steady-state
approximation. Further, the Langevin-like coarse-grained simulations,
which accounted for the first four cumulants of the reaction events
distribution, as in {\em Methods: Simulations with near-Gaussian
  distributions}, produce results nearly indistinguishable from the
analytical expressions, and, again, at most two per cent different
from the Gillespie runs. Yet coarse-grained simulations require only
1/40th the time of their Gillespie analogue since the time step is
large, $\delta t =35$.

It is hard to imagine a practical situation in modern molecular
biology where the kinetic parameters are known well enough so that the
few per cent discrepancy between the full and the coarse-grained
simulations matters. Yet the reduction of the simulation time by the
factor of over 40 is certainly a tangible improvement.

{\em The Michaelis-Menten enzyme on a membrane: } As the next step, we
compare the algorithms when the MM enzyme is embedded in the membrane,
and random substrate-membrane binding/unbinding events happen in
addition to the MM product production [i.e., the coarse-graining is
stopped after Step 1, Fig.~\ref{reduction}(b)]. We use parameters
$k_1=0.02$, $k_{-1}=2$, $k_2=1$, $q=0.01$, and $q_0 S_{\rm
  B}=1.5$. Total time of the evolution is $T=1000$, and the initial
number of the substrates on the membrane is $S_{\rm M}(t=0)=120$.
Finally, the relaxation time of a typical fluctuation of $S_{\rm M}$
can be estimated as $\tau_{\rm M} \sim 1/[q+(\partial k_{\rm
  MM}/\partial S_{\rm M})] \sim 80$, where $k_{\rm MM}$ is the rate of
the Michaelis-Menten reaction for a given $S_{\rm M}$.

On time scales of $\sim\delta t$, the binding/unbinding events are
Poisson distributed and can be simulated by the standard
$\tau$-leaping techniques \cite{tau-leapin1}.  However, for
consistency, we simulate them similarly to the MM-reaction,
approximating the Poisson distribution by its Gram-Charlier-series.

Since, in this setup, many events are futile membrane
bindings-unbinding, the Gillespie simulations become quite time
consuming, and we only achieve convergence of the first three
cumulants in a reasonable time, with the third cumulant known to about
two significant digits. For the coarse-grained runs, we choose the
time step $\delta t=20 \ll \tau_{\rm M}$, and we model all three
coarse-grained reactions preserving their first three cumulants only.
In this example, our coarse-grained approach speeds simulations
60-fold, yet it still provides accurate results for the first three
cumulants of the distribution, see Tbl.~\ref{tbl_cg_time}.

\begin{table}[t]
\begin{tabular}{|c|ccc|c|}
  \hline
  cumulant &  Gillespie    &  Coarse-grained (step1) & Coarse-grained (step 2) &  Analytical prediction  \\
  \hline
  $c_1$   & 418.7 $\pm$ 0.1 & 420.0 $\pm$ 0.1  & 418.9 $\pm$ 0.1 &418.9  \\
  $c_2/c_1$   & 0.771 $\pm$ 0.001 & 0.764 $\pm$ 0.002 & 0.768 $\pm$ 0.001 & 0.767 \\
  $c_3/c_1$  & 0.50 $\pm$ 0.03 & 0.46 $\pm$ 0.08 &0.48 $\pm$ 0.03& 0.472 \\
  time   & 1h 14min & 1min 17s &1s& N/A   \\
  \hline
\end{tabular}
\caption{\label{tbl_cg_time}Comparison of cumulants of the product
  flux for  the full system membrane calculated  using the Gillespie
  simulations, the coarse-grained
  simulations (with $\delta t=20$), the fully coarse-grained simulations ($\delta t =1000$),
 and the analytical predictions. The  data
  were averaged over $10^6$ realizations, sufficient to estimate the third
  cumulant to two significant digits.
}
\end{table}

{\em The full $S_{\rm B}\to P$ conversion:} Finally, we perform
similar benchmarking for the Gillespie simulations and the
coarse-grained simulations of the fully coarse-grained system,
represented as a single complex reaction $S_{\rm B} \to P$. The third
column in Tbl.~\ref{tbl_cg_time} presents the data for this
coarse-graining level. Note that representing all five reactions as a
single one results in a dramatic speedup of about 4000. This number
relates to the ratio of the slow and the fast time scales in the
problem, but also to the fact that futile bindings-unbindings are
leaped over in the coarse-grained scheme.

\subsection{Generalizations to a network of reactions}

\begin{figure}[t]
\centerline{\includegraphics[width=12 cm]{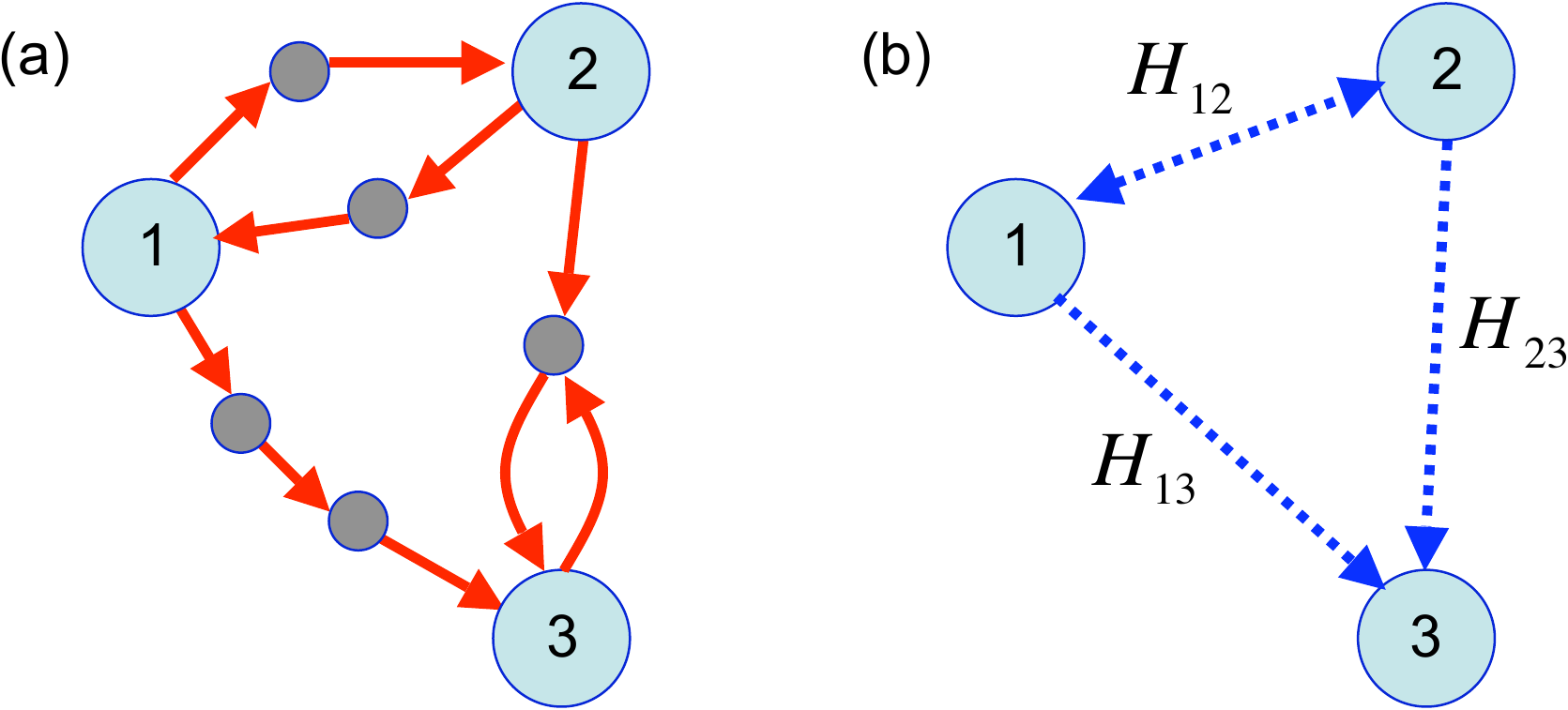}}
\caption{\label{generalization}Schematic coarse-graining of a network
  of reactions. (a) This network has $M=10$ reactions (red arrows)
  and $N=8$ species, of which three are slow (large circles), and five
  are fast (small circles). (b) Dynamics of each fast node can be
  integrated out, leaving effective, coarse-grained, pairwise fluxes
  among the slow nodes. The fluxes along entire pathways connecting
  the slow pairs (blue arrows) are labeled by the corresponding
  effective Hamiltonians $H_{\mu\nu}$. Note that, for reversible
  pathways ($H_{12}$ in our example), the flux may be positive or
  negative (two-sided arrow), and it is strictly non-negative for the
  irreversible pathways (one-sided arrows). }
\end{figure}

As discussed in detail in the original literature (the best
pedagogical exposition is in Ref.~\onlinecite{pilgram-04}), in the
stochastic path integral formalism, a network of $M$ reactions with
$N$ chemical species (cf.~Fig.~\ref{generalization}) is generally
described by $2MN$ ordinary differential equations specifying the
classical (saddle point) solution of the corresponding path integral.
{\em Methods: Coarse-graining all membrane reactions} provides a
particular example of this technique, and we refer the interested reader
the original work \cite{pilgram-04}.  Here, we build on 
the result\cite{pilgram-04} and focus on developing
a relatively simple, yet general
coarse-graining procedure for more complex reaction networks.

At intermediate time scales, $\delta t$, many fast reactions
connecting various slow variables can be considered statistically
independent. Therefore, in the path integral, every separate chain of
reactions that connects two slow variables simply adds a separate
contribution to the effective Hamiltonian. Namely, let's enumerate
slow chemical species by $\mu,\nu,\dots$. Chains of fast reactions
connecting them can be marked by pairs of indexes, e.g., $\mu\nu$
(cf.~Fig.\ref{generalization}).  An entire such chain will contribute
a single effective Hamiltonian term, $H_{\mu \nu}(\{ N \},\{ \chi
\},\{ \chi_C \})$, to the full CGF of the slow fluxes, where $\{ N
\}$, $\{ \chi \}$, and $\{ \chi_C \}$ are the set of the slow species
and the conjugate, counting variables. If necessary, the geometric
correction to the CGF, ${\mathcal S}_{\rm geom}^{\mu \nu}(\{N \},\{
\chi \},\{ \chi_{\rm C}\})$, can also be written out. Overall,
\begin{multline}
  {\mathcal S}(\{\chi_{\rm C} \},T)=\sum \limits_{\mu<\nu} {\mathcal
    S}_{\rm geom}^{\mu \nu}(\{ N(t) \},\{ \chi(t) \},\{ \chi_C \},T)\\+\int
  \limits_0^T dt \left[ \sum \limits_{\mu} i\chi_{\mu} \dot{N}_{\mu} + \sum
    \limits_{\mu<\nu} H_{\mu \nu}( \{ N(t) \},\{ \chi(t) \},\{ \chi_{\rm C}
    \})\right].
\label{netpath}
\end{multline}

This expression provides for the following coarse-graining procedure.
First, one finds a time scale $\delta t$, small enough for the slow
species to be considered as almost static, and yet fast enough for the
fast ones to equilibrate. If the fast species consist only of a few
degrees of freedom, like in the case of a single enzyme, one can
derive the CGF of the transformations mediated by these species either
by using techniques presented in this article (cf.~{\em Methods:
  Coarse-graining the Michaelis Menten reaction}), or discussed
previously \cite{nazarov-03,gopich-06,sinitsyn-07epl}. If instead the
fast species are mesoscopic, one can use the stochastic path integral
technique to derive the CGF by analogy with Step 2 of this article. 

At the next step, these expressions for the CGFs of the fast species
are incorporated into the stochastic path integral over the abundances
of the slow variables. For this, one writes down the the full
effective Hamiltonian, Eq.~(\ref{netpath}), assumes adiabatic
evolution, and solves the ensuing saddle point equations. The extremum
of the effective Hamiltonian determines the cumulant generating
function. For hierarchies of time scales, this reduction procedure is
repeated at every level of the hierarchy.

\section{Discussion}
As biology continues to undergo the transformation from a qualitative,
descriptive science to a quantitative one, it is expected that more
and more rigorous analysis techniques developed in physics, chemistry,
mathematics, and engineering will find suitable applications in the
biological domain. This article represents one such example, where
adiabatic approach, paired with the stochastic path integral formalism
of mesoscopic statistical physics, allows one to coarse-grain stochastic
biochemical kinetics systems.

For stiff systems with a pronounced separation of time scales, our
technique eliminates relatively fast variables. It reduces stochastic
networks to only the relatively slow species, coupled by complex
interactions that accounts for the decimated
nodes. The simplified system is smaller, non-stiff, and hence easier
to analyze and simulate, resulting, in particular, in orders of
magnitude improvement in the computational complexity of the
simulations. Thus we believe that the approach has a potential to
revolutionize the field of simulations in systems biology, at least
for systems with the separation of time scales.

\begin{figure}[t]
\centerline{\includegraphics[width=7 cm]{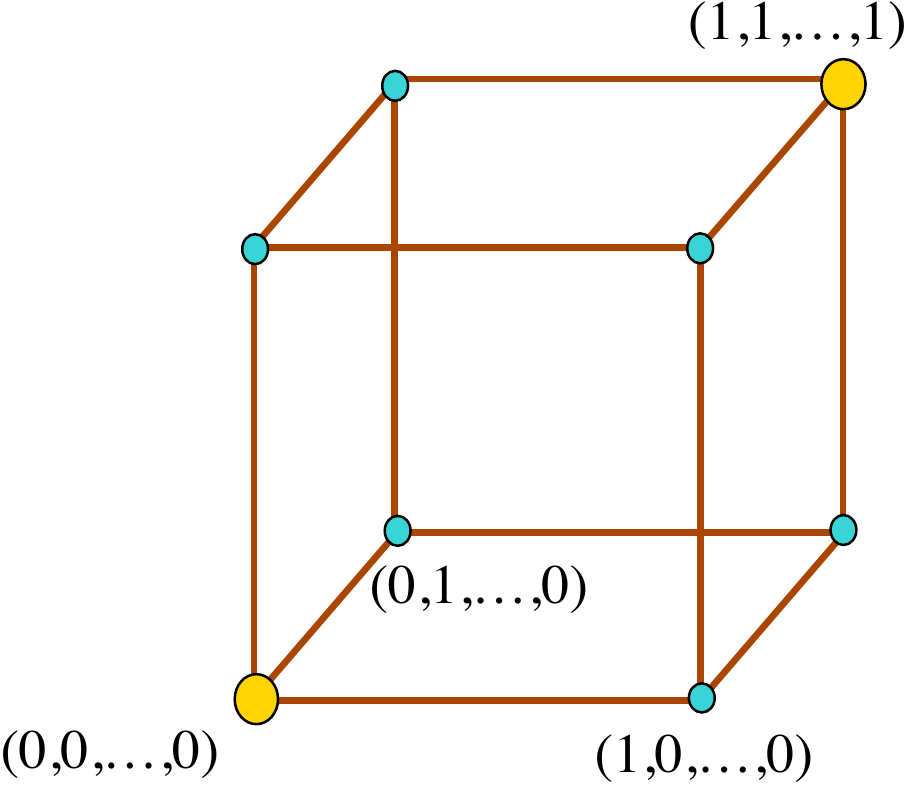}}
\caption{\label{cube}A molecule must be modified on $n$ sites (here
  $n=3$) in an arbitrary order to get activated. 0 and 1 indicate a
  non-modified/modified site, respectively. The number of states with
  $m\le n$ modified sites is quite different for different $m$'s,
  which allows for a separation of time scales, as explained in the
  text.}
\end{figure}

Fortunately, such separation occurs more prominently in Nature than
one would intuitively suspect. Consider for example, the system given
in Fig.~\ref{cube}, briefly mentioned in the {\em Introduction}. A
molecule must be modified on $n$ sites in an arbitrary order to move
from the inactive ($0,0,\ldots,0$) to the active ($1,1,\ldots,1$)
state. The kinetic diagram for this system is an $n$-dimensional
hypercube, and the number of states of the molecule with $m$ modified
sites is ${n\choose m}$. Therefore, if the total number of molecules
is $N$, then a typical $m$ times modified state will have
$N_m=N/{n\choose m}$ molecules in it. This number may be quite small,
ensuring the need for a full stochastic analysis. More importantly, it
is quite different from either $N_{m-1}$ or $N_{m+1}$, e.g.,
$N_{m}/N_{m+1}=(m+1)/(n-m)$. As we discussed at length above,
different occupancies result in the separation of time scales, and, on
practice, the adiabatic approximation works quite well when this
separation is a factor of only a few.

In addition to the analysis and simulations, our adiabatic path
integral-based coarse-graining scheme simplifies interpretation and
understanding. For example, in certain cases, the Fano factor of the
complex $S_{\rm B}\to P$ reaction, Eq.~(\ref{fano_f}), approaches
unity, suggesting a simplified, yet rigorous, interpretation with the
entire reaction replaced by a simple Poisson step.  Hence the list of
relevant, important parameters may be smaller than suggested by the 
\textit{ab initio} description of the system, 
aiding the understanding of the involved processes and decreasing the
effective number of biochemical parameters that must be measured
experimentally. Recent theoretical analysis suggests that this may be
a universal property of biochemical networks \cite{sloppy,ziv}, with
larger networks having proportionally fewer relevant parameters.  Thus
one may hope that the rigorous identification of the relevant degrees
of freedom presented here will become even more powerful as larger,
more realistic systems are considered.

We demonstrated the strength of our coarse-graining approach in
analytical calculations of the Fano factor for the model system
(relevant for single molecule experiments), and in numerical
simulations, where the method substantially decreased the
computational complexity. While impressive, this is still far from
being able to coarse-grain large, cellular scale reaction
networks. However, we believe that some important properties of our
approach suggest that it may serve as an excellent starting
point. Namely,
\begin{itemize}
\item We reduce a system of stochastic differential equations to a
  similar number of deterministic ones, which is a substantial
  simplification (see {\em Methods}).
\item Our method can operate with arbitrarily long series of moments of
  the whole probability distribution of reaction events; i.e., it
  keeps track of mesoscopic fluctuations and even of rare events
  \cite{elgart-04}.
\item The technique is very suitable for stiff systems, 
  allowing to reduce the complexity by means of standard adiabatic
  approximations, well developed in classical and quantum physics.
\item The stochastic path integral approach can deal with species that
  have copy numbers of order unity, which are ubiquitous in biological
  systems. This is not true for many other coarse-graining techniques.
\item Finally, unlike many previous approaches, the stochastic path
  integral is rigorous, can be justified mathematically, and allows
  for controlled approximations.
\end{itemize}

In the forthcoming publications, we expect to show how these
advantageous properties of the adiabatic path integral technique allow
to coarse-grain many standard small and medium-sized biochemical
networks.

\section{Methods}

\subsection{Moments generating functions for elementary chemical
reactions}
If during a time interval $\delta t$ the rate of an elementary
chemical reaction is (almost) constant, then all reaction events are
independent, and their number can be approximated as a Poisson
variable. In its turn, the CGF of a Poisson variable is
\begin{equation}
{\mathcal S}(\chi) = k (e^{i\chi}-1)\delta t,
\end{equation}
where $k$ is the Poisson rate.

In our case, Fig.~\ref{system}, two reactions satisfy these
constraints for $\tau_E\ll\delta t\ll\tau_M$: substrate binding and unbinding
to/from the membrane. Therefore, for these reactions we have:
\begin{eqnarray}
 {\mathcal S}_1(\chi)&=&q_0 S_{\rm B}(t) (e^{i\chi}-1)\delta t,
 \label{s1}\\
 {\mathcal S}_2(\chi)&=&q S_{\rm M}(t) (e^{i\chi}-1) \delta t.
\label{s2}
\end{eqnarray}

\subsection{Coarse-graining the Michaelis-Menten reaction}

Consider the $S_{\rm M}\to P$ reaction, described mathematically as in
Eq.~(\ref{MM-1}):
\begin{equation}
S_{\rm M} +E\xrightleftharpoons[k_{-1}]{k_{1}S_{\rm M}}
C\xrightarrow{k_2} E+P.
\label{MM-methods}
\end{equation}
The probabilities of transitions between bound ($P_{\rm b}$) and
unbound ($P_{\rm u}$) states of the enzyme are given by a simple two
state Markov process
\begin{equation}
\frac{d}{dt} \left[ \begin{array}{l}
P_{\rm u}\\
P_{\rm b}
\end{array} \right] = - \left[ 
\begin{array}{ll}
\,\,\,\,k_1S_{\rm M}   &  -k_{-1}-k_2\\
-k_1S_{\rm M}    &\,\,\,\, k_{-1}+k_2
\end{array} \right]
\left[ \begin{array}{l}
P_{\rm u}\\
P_{\rm b}
\end{array} \right],
\label{ev1}
\end{equation}
where $P_{\rm u}+P_{\rm b}=1$.

Lets introduce the MGF for the number of $S_{\rm M}\to P$ transitions,
\begin{equation}
 {\mathcal Z}_3(\chi,\delta t)=e^{{\mathcal S}_3(\chi)}=
 \sum_{n=0}^{\infty} P(\delta Q_3=n|\delta t)e^{in\chi},
\label{pgf1}
\end{equation}
Here $\delta Q_\mu$ stands for a {\em charge} transferred over time
$\delta t$ in a reaction $\mu$, and $\mu=3$ is the MM reaction in toy
model, Fig.~\ref{system}.  Using Eqs.~(\ref{ev1}, \ref{pgf1}), one can
show \cite{gopich-03,nazarov-03,gopich-06,sinitsyn-07prl}, that
${\mathcal Z}_3(\chi,\delta t)$ satisfies a Schr\"odinger-like
equation with a $\chi$-dependent Hamiltonian, leading to a formal
solution
\begin{equation}
{\mathcal Z}_3(\chi,\delta t)= {\bf
  1}^+\left(e^{-\hat{H}_{\rm
      MM}(\chi,t) \delta t}\right)
{\bf p}(t_0),
\label{pdf2}
\end{equation}
where  ${\bf 1}^+=(1,1)$ is the unit vector, ${\bf p}(t_0)$ is the probability vector of initial enzyme states, and
\begin{align}
\hat{H}_{\rm MM}(\chi)&=
\left[\begin{array}{cc}
k_1 N_s   & -k_{-1}-k_2 e^{i\chi} \\
-k_1 N_s  & k_{-1}+k_2
\end{array} \right].
\label{hchi}
\end{align}
The Hamiltonian, analogous to Eq.~(\ref{hchi}), can be derived for a
very wide class of kinetic schemes
\cite{nazarov-03,gopich-06,sukhorukov-07,sinitsyn-07prb}, allowing for
a relatively straightforward extension of our methods.

The solution, Eq.~(\ref{pdf2}), can be simplified considerably if the
reaction is considered in a quasi-steady state approximation, that is
$P_u$ is equilibrated at a current value of the other
parameters. This means that the time on which the
reaction is being studied, $\delta t\sim \tau_{\rm M}$, is much larger
than a characteristic time of a single enzyme turnover, $\tau_{\rm
 E}$, so we can consider $\delta t\to\infty$ in
Eq.~(\ref{pdf2}). Then only the eigenvalue $\lambda_0(\chi)$ of the
Hamiltonian $\hat{H}_{\rm MM}(\chi)$ with the smallest real part is
relevant, and
\begin{equation}
 {\mathcal Z}_3(\chi,\delta t)= e^{-\lambda_0(\chi)\delta t}. 
\label{pdf22}
\end{equation}

It is possible to incorporate a slow time dependence of the parameters
into this answer. By analogy with the quantum mechanical Berry phase
\cite{sinitsyn-07epl,sinitsyn-07prl,sinitsyn-07prb}, the lowest order
non-adiabatic correction can be expressed as a geometric phase
\begin{equation} {\mathcal Z}_3(\chi)=e^{{\mathcal S}_3(\chi)}=e^{\int_{{\bf c}}{\bf A \cdot dk}-\int dt\lambda_0(\chi,t)},
\label{pdf222}
\end{equation}
where ${\bf A} = \langle u_0(\chi)|\partial_{{\bf k}} u_0(\chi)
\rangle$, ${\bf k}$ is the vector in the parameter space, which draws
a contour ${\bf c}$ during the parameter evolution, and $\langle
u_0(\chi)|$ and $|u_0(\chi) \rangle$ are the left and the right
eigenvectors of $\hat{H}_{\rm MM}(\chi,t)$ corresponding to the
instantaneous eigenvalue $\lambda_0(\chi,t)$. The first term in
Eq.~(\ref{pdf222}) is the geometric phase, which is responsible for
various ratchet-like fluxes
\cite{sinitsyn-07epl,ohkubo-07aa,astumian-07pnas}.

After elimination of the fast degrees of freedom, the geometric phase
gives rise to magnetic field-like corrections to the evolution of the
slow variables. However, since these corrections depend on time
derivatives of the slow variables, they usually are small and can be
disregarded, unless they break some important symmetry, such as the
detailed balance \cite{sinitsyn-07prl,astumian-07pnas}, or the
leading non-geometric term is zero.  In our model, the geometric
effects are negligible when compared to the dominant contribution when 
$\tau_{\rm E}/\tau_{\rm M}\sim1/S_{\rm M}$, and we deemphasize them in most
derivations. However, we keep the geometric terms in several formal
expressions for completeness, and the reader should be able to track
its effects if desired.

Reading the value of $\lambda_0(\chi)$ from
Ref.~\onlinecite{sinitsyn-07epl}, we conclude that the number of
particles converted from $S_{\rm M}$ to $P$ over time $\delta t$,
$\tau_{\rm E}\ll \delta t\lesssim \tau_{\rm M}$ in the adiabatic (MM)
limit is described by the following CGF:
\begin{multline}
  {\mathcal S}_3(\chi,\delta t)={\mathcal S}_{\rm geom}(\chi,S_{\rm M},\dot{S}_{\rm M})+\\
  \frac{\delta t}{2} \left[ -(k_{-1}+k_2+S_{\rm
      M}k_1)+\sqrt{(k_{-1}+k_2+S_{\rm M}k_1)^2+ 4S_{\rm M}k_1k_2
      (e^{i\chi}-1)}\right].
\label{s3}
\end{multline}

\subsection{Simulations with near-Gaussian distributions}
A probability distribution $P(\delta Q)$ with known cumulants
$c_1$, $c_2$,..., can be written as a limited Gram-Charlier expansion
\cite{edgeworth}
\begin{equation}
 P(\delta Q) \approx \Psi(\delta Q,c_1,c_2) \left[1+\frac{c_3(y^3-y)}
   {6c_2^{3/2}} + \frac{c_4(y^4-6y^2+3)}{24c_2^2}+
   \frac{c_3^2(y^6-15y^4+45y^2-15)}{72c_2^3} + \cdots  \right],
\label{edgeworth}
\end{equation}
where $y=(\delta Q-c_1)/\sqrt{c_2}$ and $\Psi(\delta Q,c_1,c_2)$ is
the Gaussian density with the mean $c_1$ and the variance $c_2$.  The
leading term in the series is a standard Gaussian approximation, and
the subsequent terms correctly account for skewness, kurtosis,
etc. Note that if all cumulants scale similarly, as is true for our
near-Gaussian case, then the terms in the series become progressively
smaller, ensuring good approximations in practice.

While the Gram-Charlier expansion provides a reasonable approximation to $P$,
generation of random samples from such a non-Gaussian distribution is
still a difficult task. However, if, instead of the random numbers
{\em per se}, the goal is to calculate the expectation of some
function $f(\delta Q)$ over the distribution $P$, $\langle f(\delta
Q)\rangle_P$, then the importance sampling technique \cite{imp-sampl}
can be used. Specifically, we generate a Gaussian random number
$\delta Q$ according to $\Psi(\delta Q,c_1,c_2)$ and define its
importance factor according to its relative probability in the
reference normal distribution and the desired Gram-Charlier approximation
\begin{equation}
 \eta=\frac{P(\delta Q)}{\Psi(\delta Q,c_1,c_2)}.
\label{impfact1}
\end{equation}
After generating $N$ such random numbers $\delta Q_\nu$,
$\nu=1,\dots,N$,  we obtain the needed expectation values as
\begin{equation}
 \la f \ra_P = \frac{\sum_{\nu=1}^{N} \eta_\nu f(\delta Q_\nu)}
 {\sum_{\nu=1}^N \eta_\nu}.
\label{av}
\end{equation}
If a current random number draw represents just one reaction in a
larger reaction network, then the overall importance factor of a Monte
Carlo realization is a product of the factors for each of the random
numbers drawn within it.

\begin{figure}[t]
\centerline{\includegraphics[width=17 cm]{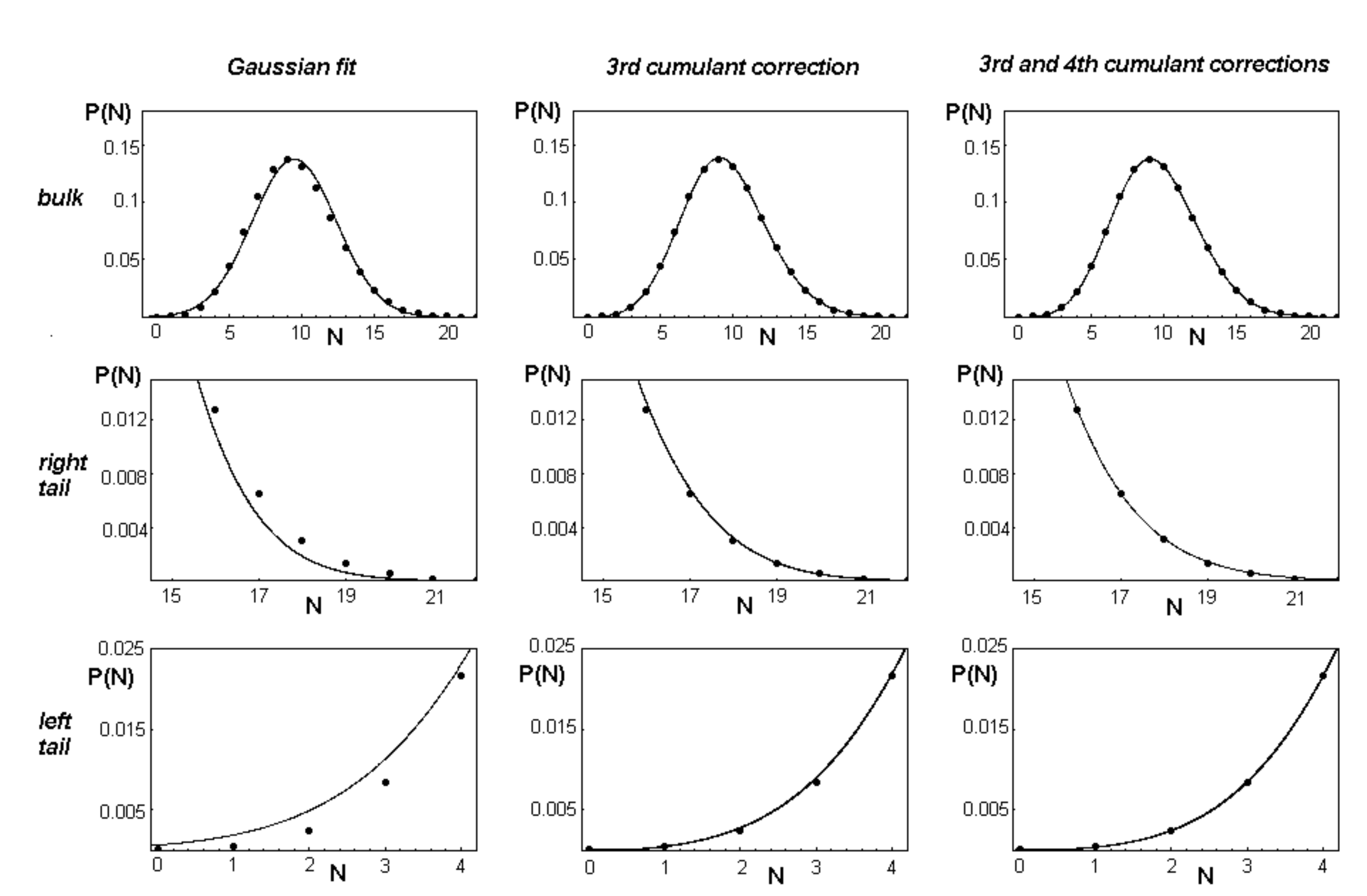}}
\caption{\label{comparison} Comparison of an exact discrete
  distribution of product molecules generated by MM-enzyme (discrete
  points), with the fit by continuous approximation by leading terms
  of Gram-Charlier series. Left column compares the exact result to the
  Gaussian approximation with the same first two cumulants. Central
  column shows improvement of the fit due to inclusion of the third
  cumulant correction.  Including the fourth cumulant (right column)
  makes the approximation and the exact result virtually
  indistinguishable. For these plots, we used $S_{\rm M}=140={\rm const}$,
  $k_1=0.02$, $k_{-1}=2.$, $k_2=1.$, $q=0.01$, and time step size
  $\delta t=35$ (see {\em Introduction: The model} and {\em Results} for
  explanation of the parameters). }
\end{figure}

Note that the method reduces the complexity of simulations to that of
a simple Gaussian, Langevin process with a small burden of (a)
evaluating an algebraic expression for the Gram-Charlier expansion, and (b)
keeping track of the importance factor for each of the Monte Carlo
runs. Yet, at least in principle, this small computational investment
allows to account for an arbitrary number of cumulants of the involved
variables. To illustrate this, in Fig.~\ref{comparison}, we compare
the Gram-Charlier-based, importance-sampling corrected simulations of the
MM reaction flux to the exact results in {\em Results: Step 1}.
Introducing just the third and the fourth cumulant makes the
simulations almost indistinguishable from the exact results.


We end this section with a note of caution: the Gram-Charlier series
produces approximations that are not necessarily positive and hence
are not, strictly speaking, probability distributions. However, the
leading Gaussian term decreases so fast that this may not matter in
practice. Indeed, in our analysis, we simply rejected any random
number that had a negative importance correction, and the agreement
with the analytical results was  still superb.

However, this simplistic solution becomes inadequate for lengthy
simulations, where the probability that one of random numbers in a
long chain of events falls into a badly approximated region of the
distribution approaches one. Then the importance factor of the entire
chain of events becomes incorrect, spoiling the convergence. In these
situations, other approaches for generating random numbers should be
used. A prominent candidate is the well-known acceptance-rejection
method \cite{rejection}. Since the true distributions we are
interested in are near-Gaussian, a Gaussian with a slightly larger
variance will be an envelope function for the Gram-Charlier approximation
to the true distribution. Then the average random number acceptance
probability will be similar to the ratio of the true and the envelope
standard deviations, and it can be made arbitrary high. Then the
rejection approach will require just a bit more than one normal and
one uniform random number to generate a single sample from the
underlying Gram-Charlier expansion. The orders-of-magnitude gain due to the
transition to the coarse-grained description should fully compensate
for this loss.  Note that, in this case, the negativity of the series
is not a problem since it will lead to an incorrect rejection of a
single, highly improbable sample, rather than an entire sampling
trajectory.

\subsection{Coarse-graining all membrane reactions}

To complete the coarse-graining step that connects
Figs.~\ref{reduction}(b) and \ref{reduction}(c), we look for the MGF
of the total number of products $Q_{\rm P}$ produced over time
$T\sim\tau_B$:
\begin{equation}
 {\mathcal Z}(\chi_C)=e^{{\mathcal S}(\chi_C)}=
 \sum_{Q_P=0}^{\infty} P(Q_{\rm P}|T)e^{iQ_{\rm P}\chi_C}.
\end{equation}
For this, we discretize the time into intervals $t_k$ of durations
$\delta t$, and we introduce random variables $\delta Q_{\mu}(t_k)$
($\mu=1,2,3$), which represent the number of each of the three different
reactions in Fig.~\ref{reduction}(b) (membrane binding, unbinding, and
MM conversion)  during each time interval. The probability
distributions of $\delta Q_\mu(t_k)$ are given by inverse Fourier
transforms of the corresponding MGFs:
\begin{equation}
 P(\delta Q_{\mu}(t_k)) = \frac{1}{2\pi} \int d \chi_{\mu} (t_k)
 e^{-i\chi_{\mu}(t_k) \delta Q_{\mu}(t_k) + H_{\mu} (\chi_{\mu}(t_k),S_{\rm B}(t_k)) \delta t},
\label{Pij}
\end{equation}
where the CGF is
\begin{equation}
 {\mathcal S}_\mu(\chi,S_{\rm B})=H_\mu(\chi,S_{\rm B})\delta t.
\label{kkk}
\end{equation}

Following \cite{pilgram-03,pilgram-04,sukhorukov-04,sinitsyn-07prl},
the MGF of the total number of product molecules created during time
interval $(0,T)$ is given by the path integral over all possible
trajectories of $\delta Q_\mu(t_k)$ and $S_{\rm M}(t_k)$:
\begin{multline}
 e^{{\mathcal S}(\chi_{\rm C},T)}=\langle e^{i\chi_{\rm C} Q_{\rm P}}\rangle =
 \int DS_{\rm M}(t_k)\prod_k \prod_\mu\int D\delta Q_{\mu}(t_k)\
 P[\delta Q_{\mu}(t_k)] e^{i\chi_C \sum_{t_k} \delta Q_{3}(t_k)}\\
 \times \delta (S_{\rm M}(t_{k+1})-S_{\rm M}(t_k) - \delta Q_{1}
 (t_k) + \delta Q_{2} (t_k)+\delta Q_3 (t_k) ).
\label{path1}
\end{multline}
Here we used the fact that $Q_{\rm P}=\sum \limits_k \delta
Q_{3}(t_k)$.

The $\delta$-function in the path integral expresses the conservation
law for the slowly changing number of substrate molecules $S_{\rm M}$.
We rewrite it as an inverse Fourier transform,
\begin{multline}
 \delta (S_{\rm M}(t_{k+1})-S_{\rm M}(t_k) - \delta Q_{1} (t_k) +
 \delta Q_{2} (t_k)+\delta Q_{3} (t_k) ) = \\
 \frac{1}{2\pi} \int_{-\pi}^{+\pi} d\chi_{\rm M}(t_k)\exp{\{
   i\chi_{\rm M}(t_k) \left[ S_{\rm M}(t_{k+1})-S_{\rm M}(t_k) -
     \delta Q_{1} (t_k) + \delta Q_{2} (t_k)+\delta Q_{3}
     (t_k)\right]\}},
\label{delf}
\end{multline}
and we substitute the expression together with Eq.~(\ref{Pij}) into
Eq.~(\ref{path1}). Then the integration over $\delta Q_{\mu}(t)$
produces new $\delta$-functions over $\chi_{\mu}$, which, in turn, are
removed by integration over $\chi_\mu(t_k)$. This leads to an
expression for the MGF:
\begin{equation}
 e^{{\mathcal S}(\chi_{\rm C},T)}=
 \int DS_{\rm M}  \int D \chi_{\rm M} \exp{\int_{0}^{T}dt [i\chi_{\rm
     M} \dot{S}_{\rm M} +
   H(S_{\rm M},\chi_{\rm M},\chi_{\rm C})]},
\label{path4}
\end{equation}
where
\begin{align}
 H(S_{\rm M},\chi_{\rm M},\chi_{\rm C})=& H_1(-\chi_{\rm M},S_{\rm
   M},t)+
 H_2(\chi_{\rm M},S_{\rm M},t)+H_3(\chi_{\rm M}+\chi_{\rm C},S_{\rm M},t) \\
 =&q_0 S_{\rm B}  (e^{-i\chi_{\rm M}}-1) + S_{\rm M}q  (e^{i\chi_{\rm M}}-1) +\nonumber\\
 &\frac{1}{2}\left[ -(k_{-1}+k_2+S_{\rm
     M}k_1)+\sqrt{(k_{-1}+k_2+S_{\rm M}k_1)^2+ 4S_{\rm M}k_1k_2
     (e^{i\chi_{\rm M}+\chi_{\rm C}}-1)}\right]
\label{hhh}
\end{align}
Notice that, unlike in the original work on the stochastic path
integral \cite{pilgram-03}, which assumed all component reactions to
be Poisson, here $H_3$ is the CGF of the entire complex MM
reaction. This we read as the coefficient in front of $\delta t$ in
Eq.~(\ref{s3}), and it is clearly non-Poisson. This ability to include
subsystems with small number of degrees of freedom, such as a single
Michaelis-Menten enzyme or a stochastic gene expression, into
coarse-graining mechanism based on the the stochastic path integral
techniques opens doors to application of the method to a wide variety
of coarse-graining problems.

Since $S_{\rm M} \gg 1$, this path integral is dominated by the
classical solution of the equations motion (i.\ e., the saddle point),
which, near the steady state, are
\begin{eqnarray}
 \dot{S}_{\rm M}=0,&& \dot{\chi}_{\rm M}=0,\\
 \frac{\partial H}{\partial \chi_{\rm M}}=0,&& \frac{\partial
   H}{\partial S_{\rm M}}=0.
\label{semicl}
\end{eqnarray}

Let $\chi_{\rm cl}(\chi_{\rm C})$ and $S_{{\rm M},{\rm cl}}(\chi_{\rm 
 C})$ solve Eq (\ref{semicl}).  Then the cumulants generating
function in the quasi-steady state approximation is
\begin{equation}
 {\mathcal S}(\chi_{\rm C},T)=T\,H(S_{\rm M,cl}(\chi_{\rm C}),\chi_{\rm
   cl}(\chi_{\rm C}),\chi_{\rm C})
\label{fcs7}
\end{equation}
This formally completes the last step of the coarse-graining by
deriving the cumulant generating function for the number of complex
$S_{\rm B}\to P$ transformation over long times.

\subsection{Simulations details}
All computer simulations were performed using Fortran 90 code, on a single
processor AMD Barton 2500 (1.83 GHz), and operating system Windows 2000.

\begin{acknowledgments}
  We thank F.\ Alexander, W.\ Hlavacek, F.\ Mu, B.\ Munsky, M.\ Wall
  for useful discussions and critical reading of the manuscript.  This
  work was funded in part by DOE under Contract No.\
  DE-AC52-06NA25396.
\end{acknowledgments}

\end{document}